\documentclass{elsart4}
\usepackage{amssymb}
\usepackage{epsfig}
\usepackage{amssymb}

\begin{document}

\begin{frontmatter}


 \title{Shot noise in HTc superconductor quantum point contact system
 }
 \author{J. A. Celis G. and William J. Herrera\corauthref{cor1}
 }
 \address{Departamento de F\'{\i}sica, Universidad Nacional de Colombia, Bogot\'a, Colombia
 }
 \corauth[cor1]{Tel: 57-1-3165000 ext 13004; E-mail: jacelisg@unal.edu.co, jherreraw@unal.edu.co}





\begin{abstract}
We study the electrical transport properties of a quantum point contact between a lead and a Hight Tc superconductor. For this, we use the Hamiltonian approach and non-equilibrium Green functions of the system. The electrical current and the shot noise are calculated with this formalism. We consider $d_{x^2-y^2}$, $d_{xy}$, $d_{x^2-y^2}+is$ and $d_{xy}+is$ symmetries for the pair potential. Also we explore the $s_{+-}$ and $s_{++}$ symmetries describing the behavior of the ferropnictides superconductors.
We found that for $d_{xy}$ symmetry there is not a zero bias conductance peak and for $d+is$ symmetries there is a displacement of the transport properties. From shot noise and current, the Fano factor is calculated and we found that it takes values of effective charge between $e$ and $2e$, this is explained by the diffraction of quasiparticles in the contact. For the $s_{+-}$ and $s_{++}$ symmetries the results show that the electrical current and the shot noise depend on the mixing coefficient, furthermore the effective electric charge can take values between $0$ and $2e$, in contrast with the results obtained for $s$ wave superconductors.
\end{abstract}

\end{frontmatter}

\section{Introduction}

\label{Introduction}

\begin{figure}[tbp]
\centering
\includegraphics[width=3.7cm]{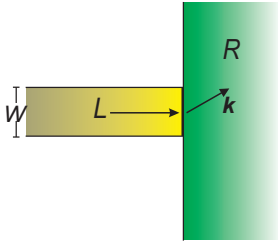}
\caption{Scheme of a normal metal lead with width $w$ (L Region) connected to a high Tc
superconductor (R Region), $\protect\theta$ is the angle of the quasiparticles in the superconductor. }
\label{Fig.1}
\end{figure}

There are two current noise sources, one is thermal fluctuations, which
causes changes in the occupation number and is known as Johnson-Nyquist
noise \cite{Blanter}. The other is the discrete behavior of the electric
charge and is known as shot noise \cite{Beenakker}. From measures of shot
noise ($S$) and electrical current ($I$) can be obtained information that
usually is not found in conductance measures as the effective electric
charge, by means of the ratio between $S$ and $I$ called Fano factor ($F$) 
\cite{Texier}. The shot noise in superconducting systems has attracted
attention due to the possibility to obtain effective electric charge equal
to $2e$, this has been analyzed for plain junctions and non-isotropic
superconductors \cite{Tanaka,Lofwander,Choi}. Nevertheless, the shot noise has not been studied in a high Tc superconductor connected to a quantum point contact, where the differential
conductance and the other properties are affected by the size of the
electrode. Some studies had shown an isotropic component in the pair
potential of HTcS \cite{Kirtley,Yong,Biswas}, this isotropic compound affect
the magnitude and phase of the pair potential and the transport properties.
Additionally, the recent discovery of high-Tc superconductivity in
ferropnictides \cite{Kamihara} has stimulated experimental and theoretical
studies of these new superconductors. The main feature of these systems is
their multiple band structure near of the Fermi level. In a simplified
scheme, the structure is reduced to two band model, where the symmetry of
the pair potential in each band could be different, and experimental evidence
has been favorable to the $s_{++}$ or $s_{+-}$ symmetries \cite{Kamihara,Golubov,Yan-Yang}.

In the present work we show the results of the electrical current, the shot
noise and the Fano factor  for a superconducting point contact system using
the hamiltonian approach. We analyze these results considering $s$, $d$, $%
d+is$ and $s_{+-} $ symmetries for the pair potential.

\section{Theory}

One method used to determine the transport properties in superconducting
junctions, is the Hamiltonian approach \cite{Bardeen}. The main idea of this
approach is to write a Hamiltonian that describes every region involved in
the junction and a coupling term between those regions, which  gives
information about the probability of one particle crossing from one region
to other. The Hamiltonian approach allows us to determine the transport
properties using the non-equilibrium Green functions of the system \cite%
{Cuevas}. We consider a semi-infinite superconductor ($R$-region) connected
to one normal metal lead ($L$ region). Our aim is to find the electrical
current, the shot noise and the Fano factor when a voltage $V$ is applied
between the lead and the superconducting region. For $d$ wave
superconductors, we consider two cases: the first $d_{x^{2}-y^{2}}$ symmetry
where $\Delta _{x^{2}-y^{2}}(\theta )=\Delta _{0}\cos {(2\theta )}$ and $%
d_{xy}$ where $\Delta _{xy}(\theta )=\Delta _{0}\sin {(2\theta )}$, with $%
\theta =\sin ^{-1}\left( k_{y}/k_{F}\right) ,$ $k_{y}$ the wavenumber in the 
$y$ direction and $k_{F}$ the wavenumber at the Fermi level. For mixed
symmetries, we include an isotropic component in the pair potential for $%
d_{x^{2}-y^{2}}$ and $d_{xy}$ wave, so that the pair potential is written as 
$\Delta _{d+is}(\theta )=\Delta _{d}+i\Delta _{s}$, where we have used $%
\Delta _{s}=0.05\Delta _{0}$ \cite{Kirtley}. To describe the transport properties of
ferropnictides we consider two models $s_{++}$ and $s_{+-}$ \cite{Golubov},
in the first we consider that the phase difference between the two gap is $0$
and for the other is $\pi $, $\Delta _{1}/\Delta _{2}=\pm \left\vert \Delta
_{1}\right\vert /\left\vert \Delta _{2}\right\vert ,$ where $\Delta
_{1\left( 2\right) }$ is the pair potential in the band $1(2)$.

Using the Keldysh formalism, the electrical current can be written in terms
of the non-equilibrium Green functions of the system $\hat{G}_{ij,nn^{\prime
}}^{\gamma \delta }$ as \cite{Cuevas}

\begin{equation}
I=\frac{e}{h}t\int_{-\infty }^{\infty }dETr\left( \hat g_{LL}^{+-}\hat\sigma
_{z}\hat G_{RR}^{-+}-\hat g_{LL}^{-+}\hat\sigma _{z}\hat G_{RR}^{+-}\right) ,
\label{FinalI}
\end{equation}

\noindent where $t$ is a parameter related with the transmission coefficient
for a lead connected to a semi-infinite normal metal ($\Delta =0$) as $T_{N}=%
\frac{4t^{2}}{(1+t^{2})^{2}}$, the superscripts $+-$ and $-+$ are the
branches in the Keldysh space, $\hat{g}_{LL(RR)}$ is the Green function of
the lead (superconductor) without perturbation ($t=0$) in Nambu space and $%
\sigma _{z}$ is the Pauli matrix. The functions $G^{+-(-+)}$ are calculated
from the Green functions of the system perturbed $\hat{G}^{r}$ solving
the Dyson equation \cite{Cuevas}. The symbol $\hat{}$ denotes $2\times 2$ matrices in
Nambu space.

In order to calculate the shot noise, we use the spectral density of the
electrical current fluctuations, and taking into account the current
expression, we obtain a similar equation for the shot noise (\ref{shot noise}). At zero 
temperature, non zero voltage and zero frequency, ($T=0$, $V\neq 0$ 
 and $\omega =0$) the shot noise is

\begin{equation}
S\left( 0\right) =\frac{4e^{2}t^{2}}{h}\int dE Tr\left[ \hat G_{LL}^{+-}\hat
G_{RR}^{-+} + \hat G_{LR}^{+-}\hat G_{LR}^{-+}\right] ,  \label{shot noise}
\end{equation}

Finally, the Fano factor is defined as the ratio between the shot noise at
zero frequency and the electrical current $F=\frac{S}{2eI}$. In the
tunneling limit the Fano Factor ($F$) gives information about the effective
charge.

\section{Results}

To calculate $\hat{G}_{ij}$ and $\hat{g}_{ij}$ ($i,j=R(L)$) we use the
superconductor surface Green function for HTc superconductors $\hat{g}%
_{S}^{r}(E,k_{y})$ \cite{Williamdos} with $E$ the energy of the quasiparticle.
From $\hat{g}_{S}^{r}\left(E,k_{y}\right) $ we obtain the Green function for a one dimensional
electrode coupled to a superconductor by

\[
\hat{g}_{RR}^{r}\left( E\right) =\sum_{k_{y}}|p\left( k_{y}\right) |^{2}\hat{%
g}_{S}^{r}\left( E,k_{y}\right) .
\]

\noindent this Green function ($g_{RR}^{r}$) is affected by the anisotropy
of the superconductor pair potential and depends on the thickness for the
contact. As we consider a contact that has only one transmission mode, the
thickness of the lead satisfies the ratio $\frac{\pi }{k_{F}}<L<\frac{2\pi }{%
k_{F}}$ and then $p\left( k_{y}\right) $ is proportional to the wave number
on $x$ direction ($k_{0xF}$) and to the overlapping between the wave
functions on $y$ direction, $p(k_{y})=k_{0xF}\left\vert \left\langle
k_{1}|k_{y}\right\rangle \right\vert ^{2}$ \cite{William_nano}.

In this work we have fixed $\Delta_0=20 meV$ and the ratio $\Delta
_{0}/E_{F}\sim10^{-1}$, which are typical values for a HTcS.

For voltages $eV<\Delta_0$, general expressions that describes the
contribution to the electrical current and the shot noise due to Andreev
reflections (AR) are

\begin{eqnarray}
I_{A} &=&\frac{16e}{h}t^{4}\int_{0}^{eV}dE\frac{|g_{RR,12}(E)|^{2}}{%
|D_{2}(E)|^{2}},   \label{ec:fano_puntual} \\
S_{A} &=&\frac{64\pi ^{2}e^{2}t^{4}}{h}\int_{0}^{eV}dE\frac{%
|Im[g_{RR,12}(E)(1+t^{4}D_{1}(E))]|^{2}}{|D_{2}(E)|^{2}}. \nonumber 
\end{eqnarray}

\noindent Therefore, in this case  the Fano factor is given by 
\begin{equation}
F=2e\frac{{\int_{0}^{eV}dE|Im[g_{RR,12}(E)(1+t^{4}D_{1}(E))]|^{2}}}{{%
\int_{0}^{eV}dE|g_{RR,12}(E)|^{2}}}.
\end{equation}

\noindent where $D_{1}=$det$\left( \hat{g}_{RR,11}^{r}\right) $ and 

\noindent$
D_{2}=1+it^{2}(\hat{g}_{RR,11}^{r}+\hat{g}_{RR,22}^{r}-t^{4}D_{1}).
$

\subsection{$s$ wave superconductors}

\begin{figure}[h]
\includegraphics[width=6.8cm]{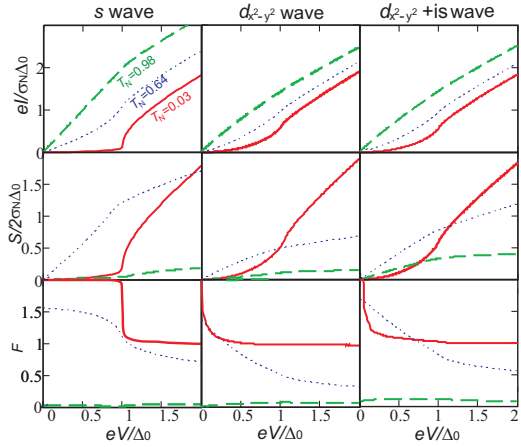}
\caption{The electrical current, the shot noise and the Fano factor (First,
second and third row) for $s$, $d_{x^2-y^2}$ and $d_{x^2-y^2}+is$ symmetries
(First, second and third column) as a function of the voltage for different
values of the transmission $T_N$.}
\label{Fig.2}
\end{figure}

For this symmetry, we consider $\Delta =\Delta _{0}$, which means that the
magnitude of the pair potential does not depend on $k_{y}$. In this case, in
the tunneling limit ($t\rightarrow 0$), the shot noise and the electrical
current are due only to AR and $F$ is equal to $2$ for voltages less than
the energy gap. For voltages higher than $\Delta _{0}$ the quasiparticles
transmission increases and the $F$ value decays to $1$ (Fig. \ref{Fig.2}).
This behavior is because the AR are equal to $1$ for $eV<\Delta _{0}$ and
tend to zero for $eV>\Delta _{0}$, this result is similar to the obtained
for plain junctions \cite{Tanaka}.

\subsection{$d$ wave superconductors}

\begin{figure}[h]
\includegraphics[width=6.3cm]{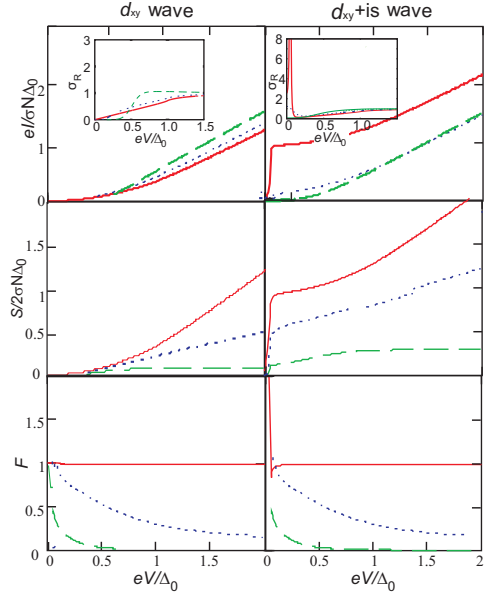}
\caption{The electrical current, the shot noise and Fano factor for $d_{xy}$
and $d_{xy}+is$ symmetries as function of the voltage for different
transmission values $T_N$. The insets show the differential conductance where we can see a peak at $eV=\Delta_s$ for $d_{xy}+is$ symmetry.}
\label{Fig.3}
\end{figure}

For $d_{x^{2}-y^{2}}$, the $F$ behavior shows that the effective charge
tends to $2e$ only when $eV$ tends to zero, that is because at zero voltage $%
eV<\Delta (\theta )$ for every $\theta $ and the Andreev reflection is the
mechanism that transport charge to the superconductor. For $eV>0$, the quasiparticle probability transmission is not
zero for $eV<\Delta _{0}$ due to
the potential anisotropy and it produces a reduction on the effective
charge and the Fano factor decreases. For $eV\gg\Delta _{0}$, the electric
effective charge tends to $e$. For $d_{xy}$ symmetry the Andreev reflection
is suppressed and it does not exist effective charge related with AR.
Therefore the electrical current and the shot noise are produced only by
quasiparticle transmission, for this reason the $F$ is equal to $1$ for
every voltage (Fig. \ref{Fig.3}). This result contrasts with the one
obtained for a plain junction \cite{Tanaka}, where the Fano factor
increases from $0$ to $1$ when the voltage is increased.

For mixed symmetries we have found that the effect of $\Delta _{s}$ in $F$.
For $d_{x^{2}-y^{2}}+is$ symmetry, $F=2$ for $eV<\Delta _{s}$, for higher
voltages $F$ decays rapidly and the behavior is the same as the
non-isotropic pair potential. For this symmetry, the electrical current and
the shot noise are similar to the properties for $d_{x^{2}-y^{2}}$ symmetry,
the difference could only be seen at low voltages (Fig. \ref{Fig.2}). For $%
d_{xy}+is$ symmetry we have found that it appears a new contribution to the
current and shot noise due to AR, this can be seen on $F$ because at low
voltages the effective electric charge is $2e$ and for higher voltages the $%
d_{xy}$ symmetry behavior is recovered. 

\subsection{Multiband superconductors, $s_{++}$ and $s_{+-}$ symmetries}

To calculate the Green functions for a lead connected to a multiband
superconductor, we mix the Green functions for two $s$ wave superconductors
by mean of a mixing coefficient $\alpha $ defined as the ratio of
probability amplitudes for an electron crossing the interface from the left
to tunnel into the first or second band on the right,

\begin{equation}
\hat{g}_{RR}^{r}(E)=\sum_{k_{y}}|t(k_{y})|^{2}\left[ \hat{g}_{\Delta
_{1}}^{r}(E,k_{y})+\alpha \hat{g}_{\Delta _{2}}^{r}(E,k_{y})\right] . 
\end{equation}

\begin{figure}[h]
\includegraphics[width=6.4cm]{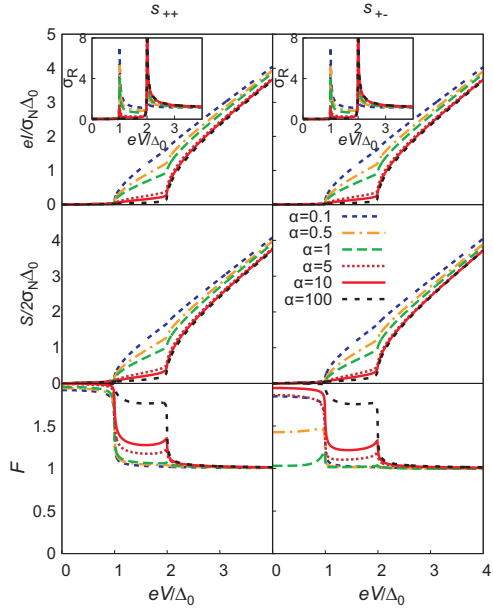}
\caption{The electrical current, the shot noise and the Fano factor for $s_{+-}$
and $s_{++}$ symmetries as function of the voltage at tunneling limit for different 
$\protect\alpha$ values. The insets show the differential conductance.}
\label{Fig.4}
\end{figure}

In figure \ref{Fig.4}, the electrical current, the shot noise and the Fano factor are shown for different $\alpha$ values in tunnel limit, we have taken $|\Delta _{1}|=2|\Delta
_{2}|=\Delta _{0}$. For $s_{++}$ symmetry when $\alpha =0.1$ the
pair potential dominant is $\Delta _{1},$ where the Fano factor is
approximately 2 for $eV\leq \Delta _{0}$ and decays to $1$ for $eV>\Delta
_{0}$, when $\alpha $ increases, there is Andreev reflection due to $\Delta
_{2}$ and the Fano factor for $\Delta_{0}<eV\leq 2\Delta _{0}$ takes values between 1 and 2 . 
In $s_{+-}$ symmetry, the pair potential $\Delta_2$ sign produces destructive interference, decreasing the Andreev reflections. The main effect is observed when $%
\alpha \simeq 1$ where the effective electric charge tends to $e$, it means
that the diffraction on the contact for this $\alpha $ value suppresses the
Andreev reflections and the electrical current is due to quasiparticles
transmission for $\Delta _{0}<eV\leq 2\Delta _{0}$.

\section{Conclusions}

We have determined the electrical current, shot noise and Fano factor for a
superconductor quantum point contact system using the Green functions and
the Keldysh formalism. We have considered five types of symmetries for the
pair potential and we found that for $d_{xy}$ symmetry, the contribution to
the current and shot noise due to Andreev reflections is null and the
effective charge for this symmetry is $e$ at every voltage. In $d_{xy}+is$
symmetry, the AR are recovered and we found that at voltages lower than $%
\Delta _{s}$ the contribution to the current increases and the effective
electric charge is $2e$. At higher voltages, the current and shot noise
recover the behavior of the unmixed symmetries. In $d_{x^{2}-y^{2}}+is$
symmetry, the isotropic component produces a constant value of AR$=1$ for
voltages lower than the isotropic gap, which means that the Fano factor is $2
$ for these voltages. For higher voltages the effective electric charge
tends to $e$. On the other hand, for $%
s_{++}$ and $s_{+-}$ symmetries the Fano factor is affected by the value of
the pair potential $\Delta _{1}$ and $\Delta _{2}$, the relative phase
between them and the coefficient $\alpha $ that mixes. These results suggest
that the Fano factor could be used to find the symmetry in high Tc
Superconductors.

\section{Acknowledgment}

The authors have received support from COLCIENCIAS (110152128235). 




\end{document}